# Sniffing Multi-hop Multi-channel Wireless Sensor Networks

Jelena Kovač, Jovan Crnogorac, Enis Kočan, *Member, IEEE*, and Mališa Vučinić, *Member, IEEE*

*Abstract* — As wireless sensor networks grow larger, more complex and their role more significant, it becomes necessary to have an insight into the network traffic. For this purpose, sniffers play an irreplaceable role. Since a sniffer is a device of limited range, to cover a multi-hop network it is necessary to consider the deployment of multiple sniffers. This motivates the research on the optimal number and position of sniffers in the network. We present a solution based on a minimal dominant set from graph theory. We evaluate the proposed solution and implement it as an extension of the 6TiSCH simulator. Our solution assumes a 50-nodes scenario, deployed in 2x2 km outdoor area, with 10% of packet drops over all channels, when 10 sniffers are used.

## I. INTRODUCTION

Wireless Sensor Networks (WSNs) can be found in various industrial and commercial applications due to their flexibility, programmability and low-price deployments. To detect the inadvertent effects that can occur due to the distributed functionality, limited communication, computation and memory resources, it is necessary to monitor the network traffic.

Sniffers are devices with specialized software that can capture and analyze network traffic. Sniffers enable traffic analysis and performance monitoring. This is important for debugging purposes and the operation analysis of WSNs, the development and testing of novel standards, protocols and implementations in real-world applications. With sniffers, information about the operation of the network can be obtained. This includes topology discovery, reboot events of nodes, isolated nodes, routing loops, packet loss, and network latency [1].

Traditional WSNs operate at a single radio channel. To sniff the traffic in the network, it is necessary to deploy sniffers at locations covered by the network, at the used radio channel. However, some WSNs, spread the network traffic over all the available channels. One such example is the 6TiSCH stack.

6TiSCH combines the industrial performance of IEEE802.15.4 in time-slotted channel hopping (TSCH) mode, with the upper-layer stack for IoT devices defined by the Internet Engineering Task Force (IETF). The 6TiSCH physical layer is typically used in the license-free 2.4 GHz ISM band. IEEE802.15.4 splits this band into 16 channels. TSCH mode enables the multi-hop operations, and is able to cope efficiently with external interference and multipath fading channels [2].

Before introducing a sniffer into such a multi-hop, multi-channel deployment, it is necessary to consider its position and coverage area. Due to the limited sensitivity of a sniffer's receiver, a single device cannot cover a WSN that spans multiple radio hops. If arbitrary large number of sniffers are deployed, they will detect almost all packets, but this leads to unnecessary additional hardware and maintenance costs. The optimal solution in terms of cost is to place the minimal number of sniffers that receive the entirety of the network traffic.

We approach this problem by assuming that: 1) sniffer devices can simultaneously receive at all 16 channels (e.g. see [3]); 2) the connectivity matrix of the network is known. Each cell in the connectivity matrix represents a link between two nodes with a corresponding Packet Delivery Ratio (PDR). Our contribution consists of an algorithm that uses graph theory to determine sniffer locations based on the connectivity matrix of the network. We consider the existing locations of network nodes as possible locations for the sniffers. We evaluate the solution using the 6TiSCH simulator and demonstrate that it is possible to achieve 90% of packet detection with around 10 sniffers in a WSN having 50 nodes randomly distributed across 2x2 km outdoor area.

The paper is organized as follows. Section II provides an overview of the existing literature on the topic. Section III describes the proposed solution. The analysis of the obtained results is presented in Section IV. Section V gives concluding remarks.

## II. RELATED WORK

Location of sink nodes in WSNs has a great impact on energy efficiency of each node in the network. One way to conserve energy is to reduce the distance between nodes and sinks. We can conclude that this problem has many similarities with finding the optimal location for sniffers. In [4], the goal was to maximize the total path reliability between sensors and sink using Mixed Integer Linear Programing formulation (MILP) with Dijkstra's algorithm for small scale problems, and Genetic Algorithms (GAs) as a heuristic solution for large-scale problems. Authors in [5], proposed a mathematical model to determine the sink

Jelena Kovač, Faculty of Electrical Engineering, University of Montenegro, Džordža Vašingtona bb, 81000 Podgorica, Montenegro (e-mail: jelenakovac96@gmail.com )

Jovan Crnogorac, Faculty of Electrical Engineering, University of Montenegro, Džordža Vašingtona bb, 81000 Podgorica, Montenegro (e-mail: jovan.crnogorac21@gmail.com)

Enis Kočan, Faculty of Electrical Engineering, University of Montenegro, Džordža Vašingtona bb, 81000 Podgorica, Crna Gora (phone 382-20-245839, e-mail: enisk@ucg.ac.me).

Mališa Vučinić, EVA team, Inria Paris, 2 rue Simone Iff, 75012 Paris, France (e-mail: malisa.vucinic@inria.fr )

locations that minimize the average communication distances.

Through this work we aim to determine location and a number of sniffers in WSNs, having in mind cost efficiency and necessary level of the captured network traffic. The work is based on the analysis of the connectivity traces collected in a real-world deployment. Tanaka et al. [6] proposed trace-based simulation for 6TiSCH that we leverage, where simulator can yield realistic results using these connectivity traces. In our previous work called d-Argus, we proposed a software solution that can detect duplicate sniffed packets in WSN deployments [7].

### III. PROPOSED SOLUTION

When considering the placement of sniffers in a network, it is important that all nodes are in radio proximity of at least one sniffer, with a defined link quality. In this work, we assume that as an input data we have a connectivity matrix of the observed network.

To obtain a connectivity matrix, it is possible to use a tool such as Mercator [8]. Mercator enables the collection of connectivity traces on a testbed in an automated manner. The connectivity trace is a time-series dataset of link's PDR and RSSI measurements for all possible links. Mercator automatically computes the PDR and the mean RSSI values on each channel.

As 6TiSCH uses multi-channel radio propagation, it is necessary to analyze 16, potentially different connectivity matrices, one for each channel. External interference affects the quality of links on some channels. Brun-Laguna et al. [9] demonstrated that external interference from Wi-Fi is typically present in real-world IEEE802.15.4 deployments, and is also most often present in testbeds, as those are typically deployed in office buildings. Also, many other technologies use the ISM band at 2.4 GHz. Therefore, connectivity matrices on different channels differ, due to the presence of external interference.

We derive an algorithm based on the graph theory to find the node locations in the network that are convenient to be used for sniffing. To do so, we use the notion of the dominating set (see Definition 1). It is important to note that finding the dominating set of a graph is a NP-complete decision problem, and there is no efficient algorithm for determining minimal dominating set [10].

**Definition 1** A dominating set *of a graph* $G = (V, E)$ *is a subset* $V' \subseteq V$ *such that, for any* $u \in V \setminus V'$, *there exists* $v \in V'$ *such that* $(u, v) \in E$. *A dominating set* $V'$ *of G is* minimal *if no proper subset of V' is a dominating set of G*.

It is important that nodes in the network are in radio proximity of at least one sniffer with a specified link quality on all channels. We specify this link quality through the parameter called *sniffer_link_pdr*, and it is the input parameter of the algorithm. We analyze the radio propagation between nodes in the network, by using the connectivity matrices. Fig. 1 (a) shows the schematic example in the form of a graph of the connectivity matrix. A link between two nodes is denoted with the corresponding PDR value.

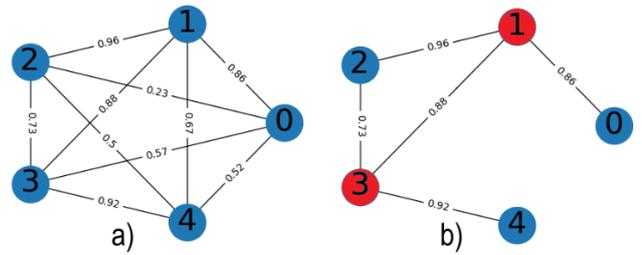

Fig. 1 (a) Schematic example of the connectivity matrix graph of the network; (b) Auxiliary connectivity matrix graph for *sniffer_link_pdr* = 0.7. Marked nodes (1 and 3) represent the minimal dominating set of the graph

We consider the existing node locations as potential locations for the sniffers. The idea is to place the sniffer as close as possible to the node at the selected location. Communication between the node and the sniffer at the same location is limited only by interference. Path loss is minimal due to the short distance. Listing Part 1 makes a simple selection of sniffers locations that covers the deployment.

---
**Part 1:** Selection of the sniffers that covers the deployment
  **Input:** Connectivity matrices Graphs $G_1...G_{16}$,
         sniffer_link_pdr
  **Output:** sniffer_candidates
1  sniffer_candidates = empty set
2  for channel in 1 .. 16 do
         // Graph $G_s$ contains links greater than sniffer_link_pdr
3    $G_s$ = make_graph($G_{channel}$, sniffer_link_pdr)
4    singleChLoc = min_dominating_set($G_s$)
5    sniffer_candidates = sniffer_candidates ∪ singleChLoc
---

In Listing Part 1, we loop through all the channels and create auxiliary connectivity graphs $G_a$ (see Fig. 1 (b)), that contain the links whose PDR values are greater than the parameter *sniffer_link_pdr*. To define sniffer locations on one channel, we generate the dominating set of the smallest size, called a *minimal dominating set*, variable *singleChLoc* (see Definition 1 and Fig. 1 (b)). Then, we repeat the same function on all 16 channels, and define union of all *singleChLoc* in the variable called *sniffer_candidates*, which is the output of the Listing Part 1. Output parameter *sniffer_candidates* contains identification (ID) of every node from the minimal dominating sets over all channels, which would be the optimal solution in case connectivity matrices on different channels are completely uncorrelated. However, because there exists a correlation among different channels, in Listing Part 2, we reduce the number of the proposed sniffer candidates.

As the input parameters for Listing Part 2 we use *sniffer_candidates* output data from Listing Part 1, and a parameter *removal_load*, number between 0 and 1. With the parameter *removal_load*, we define the percentage of sniffer candidates we attempt to remove. If the value of the parameter is set to 1, the algorithm removes as many sniffer candidates as possible.

More sniffers in the network leads to additional costs, but provides redundancy in the network, so that more packets will be detected. It is, therefore, important to find the right value of this parameter for the selected network. As mentioned before, connectivity matrices are quite similar on most of the channels, and the union solution (*sniffer_candidates*) provides sniffers that cover the same network node multiple times. Some of these sniffers can be removed without significantly degrading the coverage of the sniffers, which is the idea of Listing Part 2.

Variable *target_sniffer_num* represents the number of sniffers that we want to keep in the network. In Listing Part 2, after creating the graph $G_s$ that contains all the candidate sniffers, the algorithm attempts to remove some of the selected sniffers. The algorithm works as follows. The first sniffer from the list is removed from the graph $G_s$. Then, the algorithm checks to see if the new graph is the minimal dominating set on every auxiliary connectivity graph $(G_{a1} - G_{a16})$. If this is true, we can remove that sniffer from the selected list. Then we repeat the entire process for every selected sniffer candidate. If we reach the target sniffer number, the removal process is aborted.

```
Part 2: Try to remove some of the sniffer candidates
   Input: Auxiliary Connectivity Graphs G_a1...G_a16,
          sniffer_candidates, removal_load
   Output: reduced_candidates
   // Calculate the target number of sniffers
1  target_sniff_num =
   length(sniffer_candidates)*(1-removal_load)
   // Graph G_s contains nodes from the parameter
      sniffer_candidates
2  G_s = make_graph(sniffer_candidates)
3  sniffer_candidates = order_by(sniffer_candidates)
4  for sniffer in sniffer_candidates do
      // Remove node sniffer from graph G_s
5     G_s.remove_node(sniffer)
      // Check if the new graph is still dominating set
6     if not is_dominating_set(G_s, G_a1...G_a16) then
         // Put back node in the graph
7        G_s.add_node(sniffer)
      // Abort the removal process if we reach the target num
8     if length(G_s) ≤ target_sniff_num then
9        break
10 reduced_candidates = nodes from G_s
```

To determine whether the sniffer removal order has an influence on the results, we defined a specific sniffer removal order (see Line 3 in Listing Part 2). The idea is to remove the lowest quality sniffers first. As a quality parameter for the function *order_by()*, we chose the sum of PDR values between sniffers and other nodes in the network on all channels. These values need to be calculated for all sniffers, then sorted in descending order to remove sniffers with the worst quality.

```
Order_by: Sum of the PDR values
   Input: sniffer_candidates, all_sensors
   Output: ordered_sniffer_candidates
1  ordered_sniffer_candidates = empty set
2  for sniffer in sniffer_candidates do
3     pdr_sum[sniffer] = 0
4     for sensor in all_sensors do
         // Sum of the PDR on all channels
5        pdr_sum[sniffer] = pdr_sum[sniffer] + get_pdr(sniffer,
            sensor)
   // Sort sniffers in descending order by the PDR sum
6  ordered_sniffer_candidates = descending_order(pdr_sum)
```

## IV. EVALUATION

### A. Experimental results

In order to give an overview of the performance of the sniffer selection algorithms and to analyze their efficiency, we used the 6TiSCH simulator.

The 6TiSCH Simulator is a discrete-event simulator written in Python [11]. It captures the full behavior of the 6TiSCH stack, the Industrial IoT protocol stack standardized by the Internet Engineering Task Force (IETF). The simulator allows performance evaluation for a scenario defined with a specific set of input parameters. The simulator allows the replay of connectivity traces from testbeds or real applications to increase confidence that the results are representative of the real-world deployments [8].

We extended the 6TiSCH Simulator with a special type of a device, a sniffer, which can listen on all 16 channels, as well as with the algorithms for sniffer selection, described in Section III. Table 1 lists the parameters we used in the configuration file of the simulator. Python library NetworkX was used for the graph analysis.

We simulated networks with 50 nodes randomly deployed in the area of 2000×2000 meters, which is default simulator parameter. Each node must have at least 3 neighbors with a link PDR value greater than 0.5, on all channels. Authors in [11] have suggested these default parameters for the simulator, which ensures the generation of the networks that have properties like real-world deployments. Nodes are automatically grouped, so that the defined constraints are met. This effectively results in networks that roughly take up an area of 750×750 meters.

TABLE 1: INPUT PARAMETERS FOR SIMULATIONS

| Parameter | Value |
|---|---|
| Number of nodes | 50 |
| TSCH slotframes per run | 1000 |
| Connectivity class | Random |
| sniffer_link_pdr | 0.0,0.1,0.2,..,1.0 |
| removal_load | 0.1,0.2,0.3,..,1.0 |

We study the effect of *sniffer_link_pdr* and *removal_load* parameters, by varying them in steps of 0.1. We used the Random connectivity class of the simulator to eliminate the effect of topology on the results. Each point in the presented graphs was simulated 100 times.

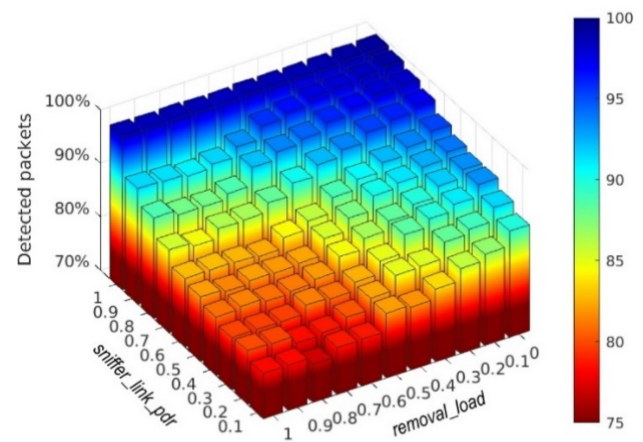

Fig. 2. Detected packet (%) at various parameters

In Fig. 2, we present the percentage of the detected packets in the network for all combinations of input parameters. As expected, by reducing the number of sniffers, fewer packets are detected. Also, by increasing the quality of the sniffer links (*sniffer_link_pdr*) the packet detection percentage increases. However, even when sniffer links are ideal (1.0), around 2% of all packets are lost. This is caused by the internal interference in the network.

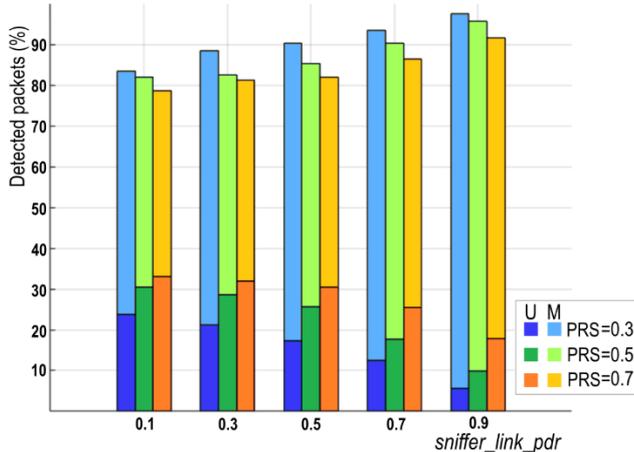

Fig. 3. Uniquely (U)/multiple (M) detected packets (%) at characteristic *sniffer_link_pdr* and *removal_load* (RL) values

Fig. 3 shows the percentage of the uniquely/multiple detected packets at characteristic parameters. Multiple detected packets refer to packets that are received by more than one sniffer. With an increase in the number of sniffers and the quality of sniffer links in the network, number of multiple detected packets increase.

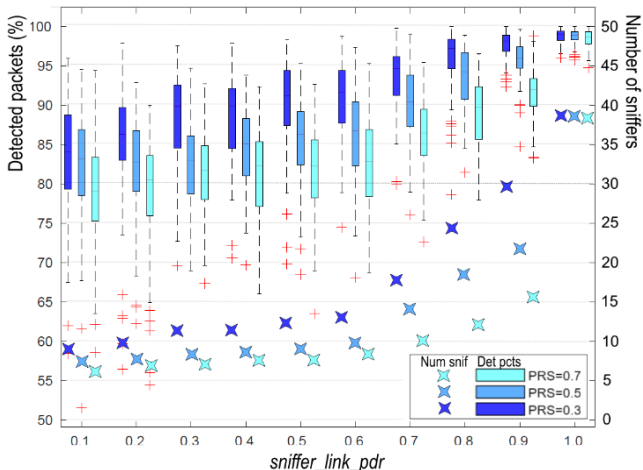

Fig. 4. Percentage of detected packet in form of box plot graph with corresponding average number of sniffers

Fig. 4 shows the percentage of the detected packets in the form of a box plot graph and the average number of sniffers at the defined parameter. We choose three values of the parameter *removal_load,* 0.3, 0.5 and 0.7, for this analysis. We can see that lower values of the sniffer link quality (*sniffer_link_pdr*) present larger variations. This occurs because the algorithm selects a different sniffer topology each time, and each of them has different quality of links, with arbitrary value between 0.1 and 1. As we increase the parameter *sniffer_link_pdr* the lower limit for the sniffer links increases and the number of different topologies that the algorithm can choose is smaller. Higher values of sniffer links' quality guarantee large percentage of the detected packets, but also lead to a large number of sniffers. It is possible to achieve around 90% of detection at the *sniffer_link_pdr* values between 0.5 and 0.8, with 10 to 15 sniffers deployed. The figure shows that with additional improvements in the selection algorithm, good results can be achieved with less than 10 sniffers, if right sniffer topology was chosen.

*B. Future work*

During the evaluation, we noticed some limitations of the simulator, namely the lack of simulated external interference. Therefore, we evaluated our solution with a module of the 6TiSCH simulator that allows replay of real-world connectivity traces. The next step is to improve the selection algorithm, using criteria that will take network interference into account. Also, we will consider the introduction of spatial analysis.

V. CONCLUSION

The paper presents the algorithm that uses the conectivity matrices and the graph theory to determine location and number of sniffers in large WSN networks. The criteria for selecting a sniffer is the sum of the PDR between the sniffer and the network nodes. The algorithm allows adjustments with the two parameters and thus allows modifications, depending on the network characteristics.